\documentclass[10pt,draftcls,letterpaper,onecolumn]{IEEEtran}
\IEEEoverridecommandlockouts
\usepackage{amssymb}
\usepackage{amsmath}
\usepackage{bm}
\usepackage{multirow}
\usepackage{algorithm}
\usepackage{algorithmic}
\usepackage{mathbbol}
\usepackage{graphicx}
\usepackage{amsthm}
\usepackage{amsfonts}
\usepackage{amsbsy}
\usepackage{amsgen}
\usepackage{amscd}
\usepackage{amstext}
\usepackage{bbold}
\usepackage{mathtools}

\begin{document}

\title{Pushing Back the Limit of \textit{Ab-initio} Quantum Transport
  Simulations on Hybrid Supercomputers}
\author{\Large Mauro Calderara$^{*}$, Sascha Br\"uck$^{*}$,  Andreas Pedersen$^*$,
  Mohammad H.~Bani-Hashemian$^{\dagger}$, \\Joost VandeVondele$^{\dagger}$,
  and Mathieu Luisier$^{*}$\\ 
$^{*}$Integrated Systems Laboratory, ETH Z\"urich, 8092 Z\"urich,
  Switzerland\\
$^{\dagger}$Nanoscale Simulations, ETH Z\"urich, 8093 Z\"urich,
  Switzerland}

\maketitle

\section*{\raggedright \Large \bf ABSTRACT}
The capabilities of CP2K, a density-functional theory package and
OMEN, a nano-device simulator, are combined to study
transport phenomena from first-principles in unprecedentedly large
nanostructures. Based on the Hamiltonian and overlap matrices
generated by CP2K for a given system, OMEN solves the Schr\"odinger
equation with open boundary conditions (OBCs) for all possible 
electron momenta and energies. To accelerate this core operation a
robust algorithm called SplitSolve has been developed. It allows to
simultaneously treat the OBCs on CPUs and the Schr\"odinger equation on
GPUs, taking advantage of hybrid nodes. Our key achievements on the
Cray-XK7 Titan are (i) a reduction in time-to-solution by more than
one order of magnitude as compared to standard methods, enabling the
simulation of structures with more than 50000 atoms, (ii) a parallel
efficiency of 97\% when scaling from 756 up to 18564 nodes, and (iii)
a sustained performance of 15 DP-PFlop/s.

\section*{\raggedright \Large \bf 1. INTRODUCTION}
The fabrication of nanostructures has considerably improved over the
last couple of years and is rapidly approaching the point
where individual atoms are reliably manipulated and assembled
according to desired patterns. There is still a long way to go before
such processes enter mass production, but exciting applications have
already been demonstrated: a low-temperature single-atom transistor
\cite{simmons}, atomically precise graphene nanoribbons \cite{fasel},
or van der Waals heterostructures based on metal-dichalcogenides
\cite{kim} were recently synthesized. 

Despite these promising advances the realization of nano-devices
remains a very tedious task, more complicated than it was at the
micrometer scale. Researchers can no longer rely on their
sole intuition and past experience to conceive properly working
nanostructures. Doing so could lead them to wrong assumptions or make
them miss relevant physical effects. Advanced technology computer
aided design (TCAD) platforms are needed to support the experimental
work and accelerate the emergence of novel device concepts. This
requires the development of accurate simulation approaches, whose key
ingredient is the bandstructure model they rely on. The latter
accounts for the material properties of the considered systems, which
might completely determine the device functionality.

\begin{figure*}[t]
\centering
\includegraphics[width=0.5\linewidth]{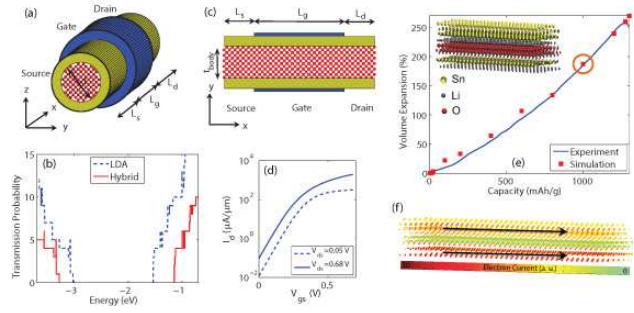}
\caption{(a) Schematic view of a 3-D gate-all-around nanowire field-effect
transistor (GAA NWFET) with a source length $L_s$, gate length $L_g$,
and drain length $L_d$, as well as a diameter $d$. Red dots indicate
atoms. The semiconducting channel is surrounded by an oxide,
represented as a uniform yellow layer. Electron transport occurs along
the $x$-axis, while $y$ and $z$ are directions of confinement. (b)
Energy-resolved transmission probability through a Si nanowire with
$d$=2.2 nm, a length $L$=34.8 nm, and composed of $N_A$=10560
atoms. Results obtained with the local density approximation (blue
lines) \cite{pade} and the HSE06 hybrid functional (red lines)
\cite{hybrid} are compared to each other. (c) Same as (a), but for a
2-D double-gate ultra-thin-body field-effect transistor (DG UTBFET) of
thickness $t_{body}$. Here, contrary to the nanowire case, the
$z$-direction (out-of-plane) is assumed periodic. (d) Transfer
characteristics $I_d$-$V_{gs}$ of a Si DG UTBFET with $t_{body}$=5 nm,
$L_s$=$L_d$=20 nm, and $L_g$=10 nm. (e) Measured \cite{wood} and
simulated \cite{pedersen} volume expansion of tin-oxide (SnO) in the
anode of a lithium-ion battery. The inset shows the atomic structure of a
lithiated SnO sample with a capacity $C$=1000 mAh/g. (f) Electronic
current through the structure depicted in the inset of (e). The two
arrows indicate the current direction. The current flow through
the central Li-oxide is insignificant.}
\label{fig:motivation}
\end{figure*}

The effective mass approximation (EMA), k$\cdot$p models \cite{kp},
tight-binding \cite{koster}, or pseudopotential methods \cite{pseudo}
offer a satisfactory level of accuracy in many applications,
but generally they suffer from several deficiencies such as their
necessary parameterization (all are empirical models), the
transferability of the parameters from bulk to nanostructures, the
treatment of heterostructures, or the absence of atomic resolution for
EMA and k$\cdot$p. \textit{Ab-initio} methods, e.g. density-functional
theory (DFT) based on the Kohn-Sham equations \cite{kohn} appear to
be more promising solutions since they address the shortcomings
mentioned above and their known band gap underestimations can be
corrected with hybrid functionals \cite{hybrid}.

First-principles codes such as VASP \cite{vasp}, ABINIT \cite{abinit},
Quantum ESPRESSO \cite{qe}, SIESTA \cite{siesta}, or CP2K \cite{cp2k}
extensively use DFT and therefore lend themselves ideally to the
calculation of electronic and crystal structures, phase diagrams,
charge densities, or vibrational frequencies in solids. However,
they are not well-suited to deal with out-of-equilibrium situations,
where, for example, an external voltage or temperature gradient is
applied to a nanostructure, inducing an electron current between
its contacts. Being able to rapidly and efficiently engineer
the magnitude and the direction of this current is a goal of utmost
importance in transistor \cite{utb}, molecular switch
\cite{mswitch}, quantum well solar cell \cite{qwsc}, quantum dot 
light-emitting diode \cite{qdled}, nanowire thermoelectric generator
\cite{hochbaum}, switching resistive memory \cite{srm}, or lithium-ion
battery \cite{chiang} research. All these devices rely on optimized
current flows that can be directly compared to measurements, contrary
to electronic structures, charge distributions, or potential profiles.

To study the transport rather than the static properties of matter,
standard DFT packages must be augmented with 
quantum transport (QT) simulation capabilities, which implies
replacing the computation of large eigenvalue problems by the
solution of linear systems of equations, as produced by the versatile 
Non-equilibrium Green's Function (NEGF) formalism \cite{datta}. While
\textit{ab-initio} tools combining DFT and NEGF have been formally
demonstrated many years ago \cite{taylor,Soler_etal02,transiesta,xue},
they remain computationally very intensive, thus limiting their
application to small systems composed of up to 1000 atoms like
molecules, nanotubes, or nanoribbons, all simulated in the ballistic
limit of transport (no scattering) \cite{app1,app2,app3,app4,app5}. 
There are two notable exceptions where the authors claim to have
reached 20000 atoms with a first-principles NEGF scheme
\cite{lmto1,lmto2}. In both studies a linearized muffin-tin-orbital
(LMTO) basis is employed, which exhibits a tight-binding-like sparsity
pattern that is mostly suitable for close-packed metallic crystals,
not for arbitrary materials, as inspected here.

To allow for the simulation of realistic nano-devices made of any
elements we propose in this paper a new \textit{ab-initio} quantum
transport solver that goes beyond the existing solutions. The
selected approach links two state-of-the-art, massively parallel
codes, CP2K \cite{cp2k} and OMEN \cite{sc11}, and leverages their DFT
and transport capabilities, respectively, in order to handle systems
composed of tens of thousands of atoms. The main target applications are
the design of nanoscale transistors and the enhancement of the
electronic conductivity in lithium ion battery electrodes, as shown in
Fig.~\ref{fig:motivation}. These are two research areas with a
tremendous need for novel simulation tools. The implementation of
the code is general enough to treat other types of nanostructures too.

OMEN already possesses advanced numerical algorithms capable of
breaking the petascale barrier while performing quantum transport
calculations, but they are optimized for tight-binding bases, not
for DFT ones, and they are restricted to CPUs only \cite{sc11}. 
Hence, a more powerful sparse linear solver called SplitSolve has
recently been developed and integrated into OMEN. It supports the usage
of CPUs and GPUs at the same time, significantly reducing the
time-to-solution of DFT+QT problems and opening the door for the
exploration of larger design spaces. With SplitSolve, the simulation
of a Si nanowire transistor composed of 55488 atoms has been
successfully achieved, which is, to the best of the
authors' knowledge, at least 10 times larger than what others
have reported so far in the literature for similar 3-D semiconducting
structures. On the Cray-XK7 Titan at Oak Ridge National Laboratory
(ORNL) we have also been able to demonstrate a sustained
double-precision performance of 15 PFlop/s in production mode.

\begin{figure}
\centering
\includegraphics[width=0.5\linewidth]{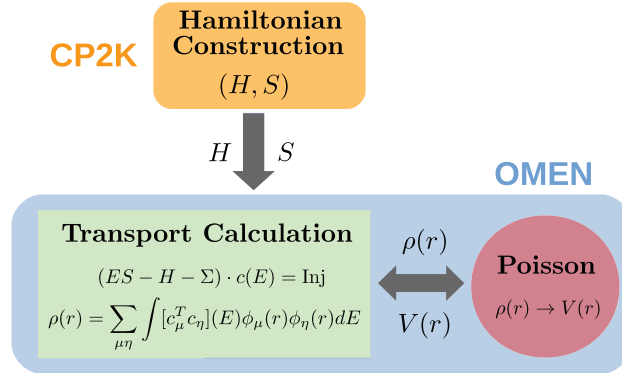}
\caption{Illustration of the OMEN+CP2K coupling scheme. For a given
  structure driven out-of-equilibrium, the Hamiltonian $H$ and overlap
  $S$ matrices produced by CP2K are transferred to OMEN, which uses
  them to solve electron transport based on the self-consistent
  solution of the Schr\"odinger and Poisson equations. More than 99\% of
  the simulation time is spent in OMEN.}
\label{fig:omen_cp2k}
\end{figure}

The paper is organized as follows: in Section 2, an
overview of the \textit{ab-initio} quantum transport approach is given,
followed by the algorithmic innovation in Section 3. A
short description of the CP2K and OMEN applications is proposed in
Section 4. The time-to-solution, scalability, and peak 
performance of the code are presented and analyzed in Section
5 before conclusion.

\section*{\raggedright \Large \bf 2. SIMULATION APPROACH}\label{sec:physics}

\subsection*{A. Density-Functional Theory with a Localized Basis}

The coupling between OMEN and CP2K is schematized in
Fig.~\ref{fig:omen_cp2k}. For a given nanostructure, CP2K starts by
solving the Kohn-Sham DFT equation \cite{kohn}
{\small
\begin{eqnarray}
\left(-\frac{\hbar^2}{2m_0}\nabla^2+V(r)+V_{H}(r)+V_{xc}(r)\right)\psi(r)=E\psi(r)
\label{eq:ks}
\end{eqnarray}
}\hspace{-0.1cm}
where the first term refers to the electron kinetic operator, the
second one, $V(r)$, to the electron-nuclei interactions, the third one,
$V_{H}(r)$, to the Hartree (Coulomb) potential, and the last
one, $V_{xc}(r)$, to the exchange-correlation energy. The wave function
$\psi(r)$ is expanded in a localized basis made of contracted Gaussian
orbitals
\begin{eqnarray}
\psi(r)&=&\sum_{\mu}c_{\mu}\phi_{\mu}(r).
\label{eq:gauss}
\end{eqnarray}
In Eq.~(\ref{eq:gauss}) $\phi_{\mu}(r)$ is a contracted Gaussian
function of type $\mu$ and $c_{\mu}$ the corresponding expansion
coefficient. Inserting this expression into the Kohn-Sham equation
Eq.~(\ref{eq:ks}) gives rise to a generalized eigenvalue problem  
\begin{eqnarray}
H\cdot c&=&E\cdot S\cdot c
\label{eq:ev}
\end{eqnarray}
with the Hamiltonian $H$ and overlap $S$ matrices as well as the
unknown expansion coefficients $c$ (now a vector) and eigenvalues
$E$. All results presented later in this paper rely on a 3SP basis
within the local density approximation (LDA) to model the exchange-correlation
energy \cite{pade}. However, the SplitSolve algorithm works with any
basis set and functional, as shown in Fig.~\ref{fig:motivation}: the
current flowing through a lithiated SnO battery anode was calculated
with a double-$\zeta$ basis and the PBE general gradient approximation
(GGA) \cite{pbe}, while the transmission probability through a Si
nanowire was obtained with the hybrid HSE06 functional \cite{hybrid}.

\begin{figure}
\centering
\includegraphics[width=0.5\linewidth]{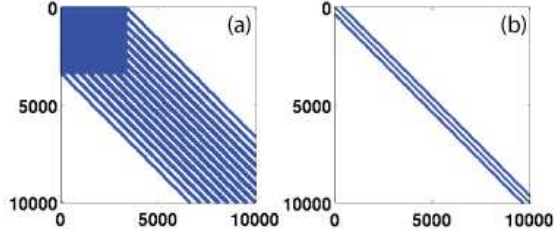}
\caption{Sub-set of the Hamiltonian matrix corresponding to a UTBFET
  as shown in Fig.~\ref{fig:motivation}(b) with $t_{body}$=5 nm. (a)
  In the contracted Gaussian basis used in CP2K. (b) In a
  tight-binding basis. The number of non-zero entries increases by two
  orders of magnitude in DFT as compared to tight-binding.}
\label{fig:ham}
\end{figure}

\subsection*{B. Quantum Transport}

\begin{figure}
\centering
\includegraphics[width=0.5\linewidth]{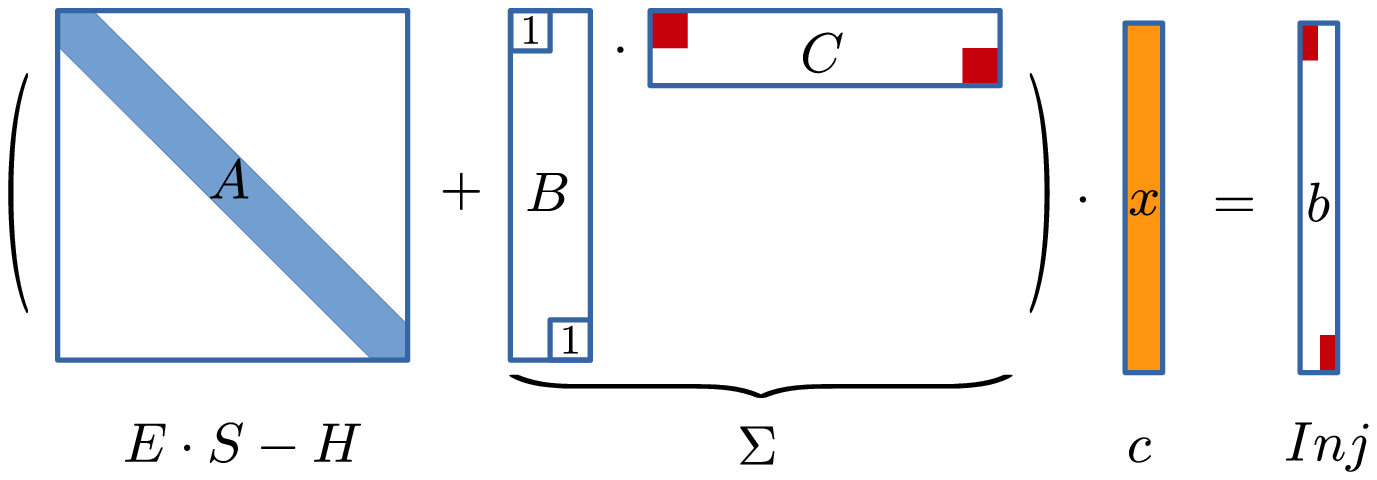}
\caption{Sparsity pattern of Eq.~(\ref{eq:wf}) and its representation
  as a $Tx=b$ sparse linear system of equations with $T=(E\cdot S-H-\Sigma^{RB})$.}
\label{fig:axb}
\end{figure}

For efficient simulations of transport through nanostructures, it is very
convenient to work within a localized basis, as the one provided by
CP2K. A sparse matrix, usually block tri-diagonal, then describes
the on-site and inter-atomic interactions. The main difference to
Eq.~(\ref{eq:ev}) is that open boundary conditions (OBCs) must be
introduced to inject electrons at the predefined contacts. These
so-called reservoirs or leads might experience different chemical
potentials, depending on the externally applied bias. This split
induces a current flow. The resulting system of equations has the
following form in the Non-equilibrium Green's Function formalism
\begin{eqnarray}
\left(E\cdot S-H-\Sigma^{RB}(E)\right)\cdot G^{R}(E)&=&\mathbb{1},
\label{eq:negf}
\end{eqnarray}
where $S$ and $H$ are directly imported from CP2K, the OBCs are
cast into the boundary self-energy $\Sigma^{RB}(E)$, $\mathbb{1}$ is the
identity matrix, and the unknowns are the retarded Green's Functions
$G^R(E)$. In the ballistic limit of transport, as here, it is computationally
more efficient to transform Eq.~(\ref{eq:negf}) into the Wave Function
formalism, which takes the form of a linear system of equations
\cite{luisier_prb_06}
\begin{eqnarray}
\left(E\cdot S-H-\Sigma^{RB}\right)\cdot c(E)&=&Inj(E).
\label{eq:wf}
\end{eqnarray}
The vector $Inj(E)$ denotes the injection mechanism into the
out-of-equilibrium devices, whereas $c(E)$ has the same meaning as in  
Eq.~(\ref{eq:ev}). The structure of Eq.~(\ref{eq:wf}) is
plotted in Fig.~\ref{fig:axb}, highlighting its sparse linear
pattern. The charge and current densities can be directly 
derived from $G^R(E)$ or $c(E)$ after Eq.~(\ref{eq:negf}) or
(\ref{eq:wf}) have been solved for all potential electron energies
$E$ and wave vectors $k$ in cases of periodicity along at 
least one direction \cite{jctn}.

The matrices $H$ and $S$ from CP2K contain about 100 times more
non-zero entries than their tight-binding or LMTO counterparts, as
shown in Fig.~\ref{fig:ham}. This is why the standard algorithms of
OMEN \cite{sc11} do not perform well in large \textit{ab-initio}
quantum transport calculations. We note that CP2K currently does not
provide any momentum- or $k$-dependence for $H$ and $S$  in periodic
systems. This issue is resolved by first cutting all the needed blocks
from 3-D simulations and then generating $H(k)$ and $S(k)$ in OMEN.

\section*{\raggedright \Large \bf 3. ALGORITHMIC INNOVATIONS}\label{sec:algo}

\subsection*{A. Open Boundary Conditions: the FEAST Algorithm}

The self-energy $\Sigma^{RB}$ and the injection vector $Inj$ in
Eq.~(\ref{eq:wf}) are calculated from the wave vectors $k_{B}$ and
eigenmodes $u_{B}$ of the leads/reservoirs of the considered
systems \cite{jctn}. The following polynomial eigenvalue problem must
be solved to obtain them  
\begin{eqnarray}
\sum_{l=-N_\mathrm{BW}}^{+N_\mathrm{BW}}{\textrm{e}^{i\cdot l\cdot
    k_{B}}(H_{q,q+l}-E\cdot S_{q,q+l})u_{B}}=0,
\label{eq:bound}
\end{eqnarray}
where $H_{q,q+l}$ and $S_{q,q+l}$ are the parts of the Hamiltonian and
overlap matrix that describe the interaction of unit cell $q$ and
$q+l$ within the leads and $N_\mathrm{BW}$ indicates the range of the
inter-cell interactions, typically $N_\mathrm{BW}\geq$2. It has been
demonstrated that in a tight-binding basis, $\Sigma^{RB}$ can be more
rapidly evaluated with Eq.~(\ref{eq:bound}) and a shift-and-invert
approach \cite{luisier_prb_06} than with the standard iterative
decimation technique used in NEGF \cite{sancho}. However, in a DFT
basis, the open boundary conditions start to represent a serious
computational bottleneck due to the size increase of the involved
matrices and the difficulty to parallelize the shift-and-invert
method.

\begin{figure}
\centering
\includegraphics[width=0.5\linewidth]{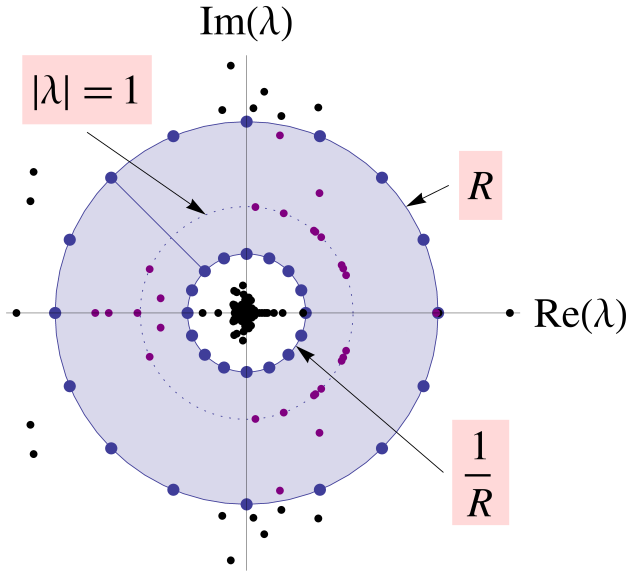}
 \caption{Selected contour in the complex plane to enclose only the
   $m$ eigenvalues corresponding to propagating and slow decaying
   modes (red dots). The black dots ($|\lambda|<$1/R and $|\lambda|>$R)
   are neglected.}
 \label{fig:feast}
\end{figure}

In practice it is not necessary to determine the $N_{BC}$ possible phase
factors $\lambda=\textrm{e}^{i\cdot k_{B}}$ as the contribution
from fast decaying modes is negligible. It suffices to find the $m$
eigenvalues inside an annulus around $|\lambda|=1$ in the complex
plane centered at the origin. Hence, a contour integration method can
be employed to find a subspace projector $Q_F$ that spans the same
space as the eigenvectors corresponding to the eigenvalues enclosed by
the shaded region in Fig.~\ref{fig:feast}. The FEAST algorithm
\cite{FEAST} and its extension to generalized eigenvalue problems with
non-Hermitian matrices \cite{LAUX} has been chosen to produce
$Q_F$. When applying the Rayleigh-Ritz method, the eigenvalue problem
reduces to a $m\times m$ system
\begin{eqnarray}
\left[Q_F^\dag A_F Q_F \right]u_B&=&\lambda \left[Q_F^\dag B_F Q_F \right]u_B
\end{eqnarray}
with the following matrix definitions
{\small
\begin{eqnarray}
A_F&=&\left(
\begin{array}{cccc}
   \tilde{H}_{-N_\mathrm{BW}+1} & \ldots & \tilde{H}_{N_\mathrm{BW}-1}  & \tilde{H}_{N_\mathrm{BW}}\\
   \mathbb{1} & & &\\
   & \ddots & &\\
   & & \mathbb{1} &\\
\end{array}
\right), 
\end{eqnarray}
\begin{eqnarray}
B_F&=&\left(
\begin{array}{cccc}
   -\tilde{H}_{-N_\mathrm{BW}} & & &\\
   & \mathbb{1} & &\\
   & & \ddots &\\
   & & & \mathbb{1} 
\end{array}
\right),
\end{eqnarray}
\begin{eqnarray}
Q_F&=&\sum_{p=1}^{N_p}{\frac{z_p}{N_p}\left(z_p B_F-A_F\right)^{-1}B_F \cdot
  Y_F}.
\label{eq:feast}
\end{eqnarray}
}
\begin{figure*}
\centering
\includegraphics[width=0.5\linewidth]{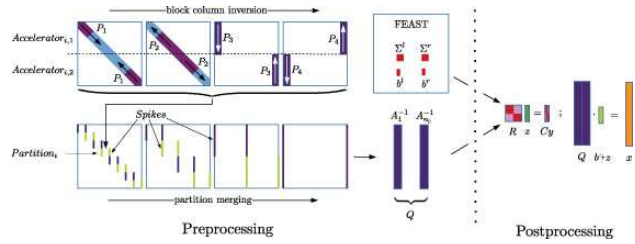}
\caption{Graphical overview of the SplitSolve algorithm on $p$
  accelerators. The system to solve is partitioned into $p$/2
  horizontal partitions. Each partition is processed by two
  accelerators with perfect parallelism. To obtain $Q = [A^{-1}_1,
    A^{-1}_{n_B}]$, the first and last block columns of $A^{-1}$, the
  partitions are merged recursively based on the SPIKE algorithm
  \cite{spike}. Upon availability of the boundary conditions
  $\Sigma^{RB}$ and the injection vectors $b$=$Inj$, the postprocessing
  phase begins. It mainly consists of solving a small system $R$ and
  performing a matrix multiplication.}
\label{fig:pp}
\end{figure*}
\noindent
Here, $\tilde{H}_j=H_{q,q+j}-E\cdot S_{q,q+j}$, the not-shown entries are
equal to 0, the size of $A_F$ and $B_F$ is $N_{BC}\times N_{BC}$, that of
$Q_F$ and $Y_F$ (matrix of random numbers) $N_{BC}\times m$ with
$N_{BC}\gg m$, and the $z_p$ are the integration points in 
the trapezoid rule. The linear systems of equations in
Eq.~(\ref{eq:feast}) determine the execution time of FEAST. Through an
analytical block LU decomposition, their size can be decreased to
$N_{BC}/(2N_\mathrm{BW})$. Furthermore, they can be solved in parallel
since the solution at each of the $N_p$ integration points is
independent from the others. FEAST can be modified according to
Ref.~\cite{beyn} to further reduce the calculation time.

\subsection*{B. Schr\"odinger Equation: the SplitSolve Algorithm}
A parallel direct sparse linear solver such as MUMPS \cite{mumps} or a
custom-made block cyclic reduction (BCR) \cite{sc11} are typically needed
to solve the Schr\"odinger equation with OBCs. Iterative solvers
cannot be efficiently used due to the presence of multiple
right-hand-sides. Since our BCR method relies on the sparsity provided
by a tight-binding basis, it does not work with DFT, whereas MUMPS
becomes slow when the number of non-zero entries in the Hamiltonian
and overlap matrices increases drastically. To address these issues we
have developed the SplitSolve algorithm that can solve
Eq.~(\ref{eq:wf}) on accelerators (GPUs or others) and leverage its
particular structure displayed in Fig.~\ref{fig:axb}. The matrix
$T=(E\cdot S-H-\Sigma^{RB})$ is of size $N_{SS}$ (total number of
atoms times orbital per atom) with a block tridiagonal shape. The
right-hand-side $Inj$ has multiple columns and non-zero elements only
in the top and bottom block rows. Consequently, solving such a system
is equivalent to obtaining the first and last block columns of
$T^{-1}$ and multiplying them with the non-zero rows of $Inj$. 

The goal of SplitSolve is twofold: (i) efficiently computing only the
required parts of $T^{-1}$ and (ii) decoupling the calculation of the
open boundary conditions $\Sigma^{RB}$ from the solution of
$T^{-1}$. This can be achieved by choosing a suitable low rank product
representation of the boundary self-energy $\Sigma^{RB}\vcentcolon=BC$, as in
Fig.~\ref{fig:axb}, and by recalling the Sherman-Morrison-Woodbury
formula     
\begin{eqnarray}
(A + BC)^{-1} = A^{-1} - A^{-1} B (\mathbb{1} + C A^{-1} B)^{-1} C A^{-1}.\nonumber
\end{eqnarray}
By doing so the computation of $\Sigma^{RB}$ can be interleaved with
the solution of the full problem. A similar approach has been
successfully tested in Refs.~\cite{cauley,parco} for
tight-binding. Defining $x\vcentcolon=c$, $b\vcentcolon=Inj$ and
introducing $Q \vcentcolon=A^{-1} B$ and $y \vcentcolon= A^{-1} b$, it
follows that
\begin{eqnarray}
x&=&T^{-1} b\quad=\quad(A-BC)^{-1} b\\
&=&y + Q(\mathbb{1} - CQ)^{-1} C y\\
&=&y + Qz,
\label{eq:x}
\end{eqnarray}
where $R \vcentcolon= (\mathbb{1} - CQ)$ and $z \vcentcolon=
R^{-1} C y$. The final solution $x$ is then obtained in four steps
\begin{itemize}
\item Step 1 Solve $AQ = B$ for $Q$.
\item Step 2 Solve $Ay = b$ for $y$.
\item Step 3 Solve $Rz=(\mathbb{1} - CQ)z=Cy$ for $z$.
\item Step 4 Compute the full solution as $x = y + Qz$.
\end{itemize}
The choice of the $B$ and $C$ matrices is free to a large extent, but
critical to minimize the computational burden. In
SplitSolve $B$ is a $N_{SS} \times 2s$ zero matrix, except the $s\times s$
top left and bottom right subblocks which are equal to $\mathbb{1}$,
while $C$ is a $2s \times N_{SS}$ matrix with the top left block set
to $\Sigma^{RB,L}$ and the bottom right one to $\Sigma^{RB,R}$, the
boundary self-energies of the left ($L$) and right ($R$)
contacts, both of size $s\times s$. Given this selection of $B$ and
$C$ the following observations are made about SplitSolve:  
\begin{itemize}
    \item The algorithm can be decomposed into a pre- and
      post-processing phase. The former consists of Step 1 and can
      be computed in parallel with the evaluation of $C$ and $b$. The
      latter comprises Steps 2, 3, and 4 and must be performed
      after the completion of the $Q$, $C$ and $b$ computation. 
    \item The structure of $A$ prior to the addition of the boundary
      conditions is preserved and can be leveraged in the
      preprocessing phase: $A$ is usually real symmetric in 3-D
      structures and complex Hermitian in 1-D and 2-D.
    \item In terms of numerics, $Q$ consists of the first and last $s$
      columns of $A^{-1}$, $y$ can be obtained as $Q \cdot b$, $R$ is
      a system of comparably small size $2s \times 2s$, and $x$ can be
      computed as $x = Q \cdot (b' + z)$ where $b'$ denotes the
      nonzero rows of $b$. 
\end{itemize}

\textbf{SplitSolve Preprocessing: } To calculate the first and last
$s$ columns of $A^{-1}$ in Step 1, a modified version of the
recursive Green's Function (RGF) algorithm \cite{rgf} has been
implemented. If $A_{i,j}$ denotes the $i^{\textrm{th}}$ block row and
$j^{\textrm{th}}$ block column of the block tridiagonal matrix $A$
with $n_B$ diagonal blocks, then the preprocessing part of SplitSolve
can be summarized as in Algorithm \ref{alg:1}.
  
\begin{algorithm}
\caption{Block column inversion on accelerators}
\begin{algorithmic}

    \STATE $X_{n_B+1} \gets 0$
    \STATE $Q_{0} \gets -\mathbb{1}$
    
    \COMMENT{Phases $P_1$ \& $P_2$ in Fig.~\ref{fig:pp}}
    \FOR{$i=n_B\rightarrow 1$}
        \STATE Solve $(A_{i,i} - A_{i, i+1} \cdot X_{i+1}) \cdot X_{i} = A_{i,i-1}$ for $X_i$
    \ENDFOR\\

    \COMMENT{Phases $P_3$ \& $P_4$ in Fig.~\ref{fig:pp}}
    \FOR{$i=1\rightarrow n_B$}
        \STATE $Q_i \gets -X_i \cdot Q_{i-1}$
    \ENDFOR \\

\end{algorithmic}
\label{alg:1}
\end{algorithm}

This gives $Q_{i,1:s}$=$A^{-1}_{i,1}$, the first block column of
$A^{-1}$. The last one, $Q_{i,s+1:2s}$=$A^{-1}_{i,n_B}$, can be
derived similarly and in parallel with $A^{-1}_{i,1}$ since the two
tasks are independent from each other and naturally scale to two accelerators. 
Alternatively, the matrix $A$ can be partitioned horizontally into two
parts of equal size and the preprocessing in SplitSolve includes four
phases, $P_1$ to $P_4$, as illustrated in Fig.~\ref{fig:pp}.

To study transport through realistic nanostructures it is crucial to
parallelize SplitSolve beyond two accelerators so that 
large system of equations can be efficiently solved. In our
implementation, the number of partitions must be a power of 2. Each
pair of accelerators computes the first and last block columns of the
inverse corresponding to its local partition using Algorithm
\ref{alg:1}. In a second step adjacent partitions are recursively
merged together based on a modified and optimized variant of the SPIKE
algorithm \cite{spike}. With the proposed scheme the merging steps
have a constant cost, their computation is evenly distributed over all
the accelerators, and their number grows logarithmically with the
number of partitions. The parallelization incurs very little memory
overhead so that the extra memory gained by the presence of additional
accelerators allows to treat bigger devices.  

\textbf{SplitSolve Postprocessing:} Upon availability of $Q$,
$\Sigma^{RB}$, and $b$, the matrices $R$ and $Cy$ are constructed on
the two accelerators storing the first and last partition and
calculating $z$. This quantity is then distributed to all the
accelerators to solve Eq.~(\ref{eq:x}) with one single
matrix-vector multiplication per block, i.e. $x = Q \cdot (b' + z)$.

\subsection*{C. Node-Level FEAST and SplitSolve Performance}
The performance of the newly implemented FEAST+SplitSolve approach has
been tested on the Cray-XC30 Piz Daint at the Swiss National
Supercomputing Centre (CSCS) \cite{daint} and on the Cray-XK7 Titan at
Oak Ridge National Laboratory (ORNL) \cite{titan}. More details about
both machines are given in Section 5A. Relevant is that
they both offer hybrid nodes with CPUs and GPUs. The main
features of FEAST+SplitSolve are then the following:
\begin{itemize}
\item The calculation of the OBCs and the solution of
  Eq.~(\ref{eq:wf}) are interleaved, FEAST being executed on the
  CPUs, while SplitSolve employs the GPUs;
\item All the operations in FEAST, e.g. the solution of
  Eq.~(\ref{eq:bound}) rely on BLAS \cite{blas} and LAPACK 
  \cite{lapack}, mostly \verb|zgemm|, \verb|zggev|, and \verb|zgesv|;
\item In SplitSolve, the matrix $A=(E\cdot S-H)$ is distributed over
  all the available GPUs and stored in their memory. Half of the
  matrix $Q$ is stored on the GPUs and half on the CPUs to reduce the
  memory consumption. The induced CPU$\leftrightarrow$GPU data
  transfer overlaps with computation (no cost);
\item The calculation of each $Q_i$ block in Alg.~\ref{alg:1}
  requires two matrix-matrix multiplications, one LU factorization,
  and one backward substitution. All the involved matrices are
  either dense or treated as such to leverage the cuBLAS
  (\verb|d/zgemm|) \cite{cublas} and MAGMA (\verb|zgesv_nopiv_gpu|)
  \cite{magma} libraries. 
\end{itemize}

\begin{figure}
\centering
\includegraphics[width=0.5\linewidth]{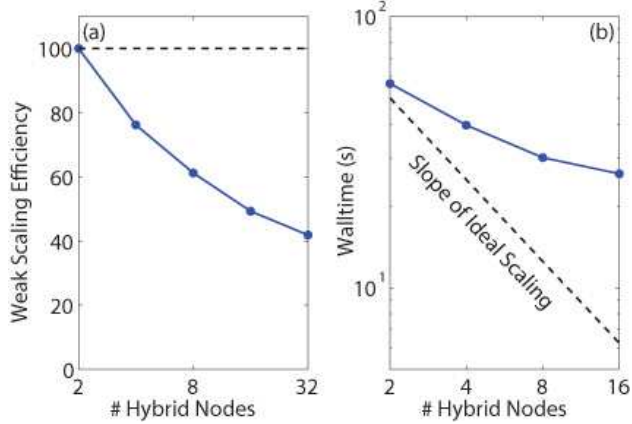}
\caption{Scaling performance of the SplitSolve algorithm on the
  Cray-XC30 Piz Daint. A Si UTBFET with $t_{body}=5$ nm is
  used as test bed. (a) Weak scaling efficiency with a constant number
  of atoms per GPU (2560, $N_{SS}=N_{GPU}\times30720$). The
  efficiency drops as the GPU count increases because of the
  additional spike calculations, as shown in Fig.~\ref{fig:pp}. (b)
  Strong scaling for a device containing 10240 atoms ($t_{body}=5$ nm,
  $L=34.8$ nm, $N_{SS}=122880$). The poor scalability is due to the
  structure size, which is the largest that 2 GPUs can handle, but
  does not offer enough workload for $\geq$8 GPUs.}
\label{fig:ss_scaling}
\end{figure}

The weak and strong scaling of FEAST+SplitSolve on Piz Daint are
reported in Fig.~\ref{fig:ss_scaling} for a 2-D ultra-thin-body
field-effect transistor (UTBFET), as in
Fig.~\ref{fig:motivation}(c). All CPUs and GPUs per hybrid node
are used. The calculation of the OBCs with FEAST is completely hidden
by the solution of Eq.~(\ref{eq:wf}). The relatively low weak scaling
efficiency originates from the extra calculation of spikes, which are
required to parallelize the workload beyond 2 GPUs. These spikes are
depicted in Fig.~\ref{fig:pp}. Their generation takes 10 sec per
recursive step so that the total simulation time increases from 30 sec
on 2 GPUs (1 partition) up to 70 sec on 32 GPUs (16 partitions, 4
recursive steps). In the strong scaling case, the limitation comes
from the impossibility to have a device structure that is large enough
to create enough work on 16 GPUs and small enough to fit onto 2 GPUs.

\begin{figure}
\centering
\includegraphics[width=0.5\linewidth]{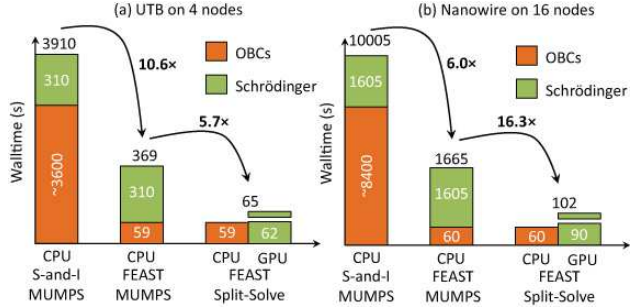}
\caption{Comparison of three different algorithms to solve the
  Schr\"odinger equation with OBCs on Titan at one given electron
  momentum and energy for (a) a Si UTBFET with $t_{body}$=5 nm,
  $L$=78.2 nm, and 23040 atoms (size of $H$/$S$: $N_{SS}$=276480) and
  (b) a Si NWFET with $d$=3.2 nm, $L$=104.3 nm, and 55488 atoms (size
  of $H$/$S$: $N_{SS}$=665856). The orange blocks refer to the time
  for the OBC calculation, the green ones to the time for the solution
  of Eq.~(\ref{eq:wf}). The first algorithm combines a shift-and-invert
  method for the OBCs \cite{luisier_prb_06} with MUMPS \cite{mumps},
  in the second one, shift-and-invert is replaced by FEAST
  \cite{FEAST}, while in the third one, SplitSolve is introduced. The
  speed-up factors between each approach are also reported.}
\label{fig:speedup}
\end{figure}

Our strategy is to choose the minimum number of GPUs that can
accommodate the desired nanostructure. This is what has been done in
Fig.~\ref{fig:speedup} on Titan where the time-to-solution of FEAST+SplitSolve
is compared to FEAST+MUMPS \cite{mumps} and shift-and-invert+MUMPS
\cite{luisier_prb_06} for a 3-D Si nanowire field-effect transistor
(NWFET) composed of 55488 atoms (16 GPUs) and  a Si UTBFET with 23040
atoms (4 GPUs). MUMPS\_5.0 has been selected because it is faster
than SuperLU\_dist \cite{superlu} for these examples. The speedup
between shift-and-invert+MUMPS and FEAST+SplitSolve is larger than 50
in both cases, i.e. our algorithms optimized for \textit{ab-initio}
transport calculations significantly outperform those designed for
tight-binding problems. Also, SplitSolve alone is between 6 and 16 
times faster than MUMPS on the same number of hybrid nodes. These
speedups demonstrate that the scaling in Fig.~\ref{fig:ss_scaling} is
not a limiting factor.

\section*{\raggedright \Large \bf 4. APPLICATION DESCRIPTION}\label{sec:appli}
The work flow of the OMEN+CP2K DFT-based quantum transport simulator is
summarized in Fig.~\ref{fig:omen_cp2k}. It is basically a
Schr\"odinger-Poisson solver with open boundary conditions to
account for electron flows between contacts. CP2K, which is a
freely available DFT package \cite{cp2kweb}, is used here to construct
the structure of the investigated nano-devices, relax their atom
positions, and generate the corresponding Hamiltonian and overlap
matrices. The latter are then transferred to OMEN, which performs
quantum transport calculations based on them. This part consumes
99\% of the total simulation time and is the focus of this work.

OMEN is a massively parallel, one-, two-, and three-dimensional
quantum transport simulator that self-consistently solves the
Schr\"odinger and Poisson equations in nanostructures. The tool was
originally based on different flavors of the nearest-neighbor
tight-binding model, but it has been extended towards
\textit{ab-initio} capabilities through its link with CP2K and the
incorporation  of the FEAST+SplitSolve approach 
presented here. It is important to realize that the OMEN version that
produced all the results discussed in the next Section is the same as
the one that is used to simulate nano-devices on a daily basis, either
in the ballistic limit of transport or in the presence of scattering
\cite{prb09}. OMEN is a real production code and not a software
that was only tuned to reach the highest possible performance on a
specific instance of a general problem. 

\begin{figure}
\centering
\includegraphics[width=0.5\linewidth]{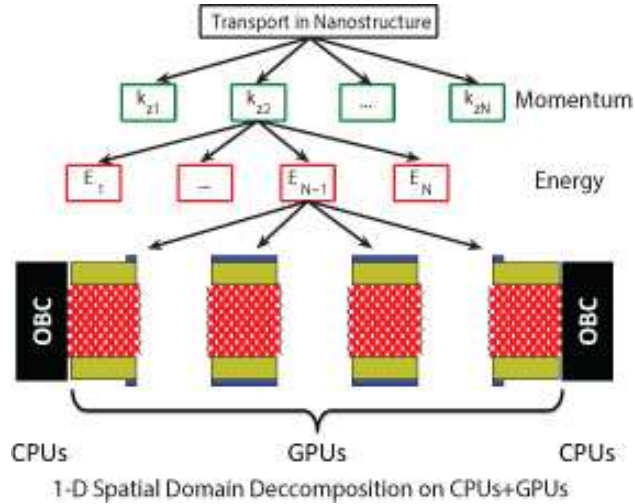}
\caption{OMEN multi-level distribution of the workload. The electron 
  momentum $k$ and energy $E$ are almost embarrassingly parallel
  and form the two highest levels. A 1-D spatial domain decomposition
  is applied to the solution of the Schr\"odinger equation with open
  boundary conditions, simultaneously stressing multiple CPUs and
  GPUs. This is the third level of parallelization.}
\label{fig:parallel}
\end{figure}

Although four natural levels of parallelism are available in OMEN,
only three have been utilized in this paper, as shown in
Fig.~\ref{fig:parallel}: the momentum $k$ and energy $E$ points are
almost embarrassingly parallel, while FEAST+SplitSolve provides a
1-D spatial domain decomposition. The distribution of the workload is
controlled with MPI \cite{mpi} and a hierarchical organization of
communicators. To avoid any work imbalance between sub-communicators
corresponding to different $k$ points, a dynamical allocation of the
number of nodes per momentum has been developed and verified before
\cite{parco}.

On the implementation side, OMEN is written in C++ and CP2K in Fortran 
2003. The coupling between the two packages currently occurs through a
transfer of binary files. Not all the nodes running OMEN load the
Hamiltonian and overlap matrices, but only those necessary to gather
all the unique parts of $H$ and $S$. The resulting data are then
distributed to all the available MPI ranks with \verb|MPI_Bcast| and
copied to the GPUs. The parts of CP2K and OMEN ported to GPUs are
implemented in the CUDA language from NVIDIA.

\begin{table}
\centering
\begin{tabular}{|c|c|c|}
\hline
&Piz Daint \cite{daint}&Titan \cite{titan}\\
\hline
Hybrid nodes&5272&18688\\
\hline
GPUs&5272&18688\\
\hline
GPU model&Tesla K20X&Tesla K20X\\
\hline
Cores&42176&299008\\
\hline
CPU model&Intel Xeon E5-2670&Opteron 6274\\
\hline
Node perf.~(GFlop/s)&166.4+1311&134.4+1311\\
\hline
\end{tabular}
\caption{Technical specifications of Piz Daint and Titan.}
\label{tab:1}
\end{table}

\section*{\raggedright \Large \bf 5. PERFORMANCE RESULTS}\label{sec:perf}

\subsection*{A. System Description}\label{sec:mach}

All the OMEN+CP2K simulations have been run on the Cray-XC30 Piz Daint
at CSCS and Cray-XK7 Titan at ORNL. The technical specifications of
both machines are summarized in Table \ref{tab:1}. The codes are
compiled with GNU 4.8.2 and CUDA 5.5/6.5 and use the Cray MPICH 7.0.4,
libsci 4.9, and MAGMA 1.6.2 libraries with customizations. On Piz
Daint, all the CPUs per node are active, while at least half of them
remain idle on Titan. The MAGMA function \verb|zgesv_nopiv_gpu| is
responsible for that: it performs a hybrid LU factorization on a CPU
core and a GPU. On Titan it has been observed that the factorization
time deteriorates if all the CPUs work because the competition between
them negatively affects the execution of \verb|zgesv_nopiv_gpu|. In
spite of that SplitSolve is still about 10\% slower per node on Titan
than on Piz Daint.

\subsection*{B. Measurement Methodology}
As can be seen in Algorithm \ref{alg:1} the number of floating point
operations (FLOPs) involved in SplitSolve is deterministic and can be
accurately estimated. Still, we decided to employ PAPI \cite{papi} to
measure the CPU FLOPs and the CUPTI library \cite{cupti} for the
GPUs. In the CPU case, the function \verb|PAPI_start_counters| with
Events=PAPI\_DP\_OPS is placed right after the loading of the
Hamiltonian and overlap matrices imported from CP2K, i.e. the input
reading time is not accounted for in the performance
measurement. Our limited access to the computational resources
of Titan is the reason for this neglection: in a full-scale run,
loading $H$ and $S$ and setting up the simulation environment takes
about 4 minutes. Then each self-consistent Schr\"odinger-Poisson
iteration lasts between 15 and 20 minutes. Knowing that an entire
simulation involves roughly 40-50 iterations for 10 bias points, the 4
initial minutes turn negligible. However, due to our limited
allocation, we could only simulate 1 or 2 iterations so that
accounting for the initialization phase is not representative of the
actual code performance. The function \verb|PAPI_stop_counters| is set
at the end of the simulation and includes the output writing time (a
couple of seconds).

On the GPUs the FLOPs are obtained by sampling the device counters
every second with the function \verb|cuptiEventGroupReadEvent|. The
accuracy of the method has been verified with examples where the
number of FLOPs is known, e.g.~matrix-matrix multiplications. This
approach cannot be used ``on-the-fly'' since it considerably slows down the
execution time. Hence, what is measured is the number of FLOPs that
arise from the solution of the OBCs on CPUs and of Eq.~(\ref{eq:wf})
on GPUs for one single representative energy point. To find the total 
FLOPs this result is then multiplied by the total number of energy
points that have been calculated.

\begin{figure}
\centering
\includegraphics[width=0.5\linewidth]{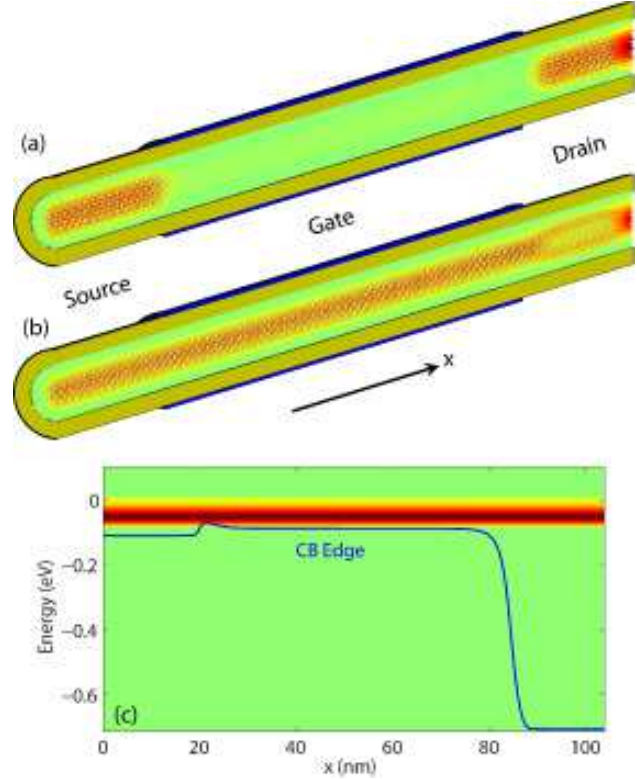}
\caption{Electron distribution (a), current map (b), and spectral
  current (c) in a Si GAA NWFET as in Fig.~\ref{fig:motivation}(a) with a
  diameter $d$=3.2 nm, $L_s$=$L_c$=20 nm, $L_g$=64.3 nm, $V_{ds}$=0.6
  V, and composed of 55488 atoms. The nanowire transport axis is $x$,
  which is aligned with the $<$100$>$ crystal axis. In all the
  sub-plots, red means high concentrations (charge or current) and
  green none. In (b) and (c) an electron current $I_d$=1.5 $\mu$A flows
  through the nanowire. In (c), the blue line indicates the conduction
  band edge of the transistor.}
\label{fig:nw}
\end{figure}

\subsection*{C. Time-to-Solution}
As explained earlier, \textit{ab-initio} quantum transport simulations
are usually limited to 1000 atoms due to their heavy computational
burden. The FEAST+SplitSolve combination makes possible the
investigation of much larger nanostructures such as a Si
gate-all-around nanowire transistor with a diameter $d$=3.2 nm, a
source and drain extension $L_s$=$L_d$=20 nm, a gate length
$L_g$=64.3 nm, and a total of 55488 atoms. A schematic view of this
device is given in Fig.~\ref{fig:motivation}(a), while its
atomically resolved charge and current distributions as well as its
spectral current are shown in Fig.~\ref{fig:nw} for one bias
point. The structure dimensions are at least one order of magnitude
larger than what is reported in the literature and comparable to what
is fabricated in laboratories \cite{suk}. As indicated in
Fig.~\ref{fig:speedup} the computational time per energy point for
this nanowire reduces to 102 sec with FEAST+SplitSolve using 16 hybrid
nodes of Titan. Hence, it has been measured that a self-consistent
Schr\"odinger-Poisson iteration including 2000 energy points takes
less than 10 minutes on 8192 nodes. Knowing that the time per energy
point with FEAST+MUMPS is in the order of 30 minutes on 16 nodes, a
CPU machine with four times as many nodes would still be 3$\times$
slower than our refined approach. 

\begin{figure}
\centering
\includegraphics[width=0.5\linewidth]{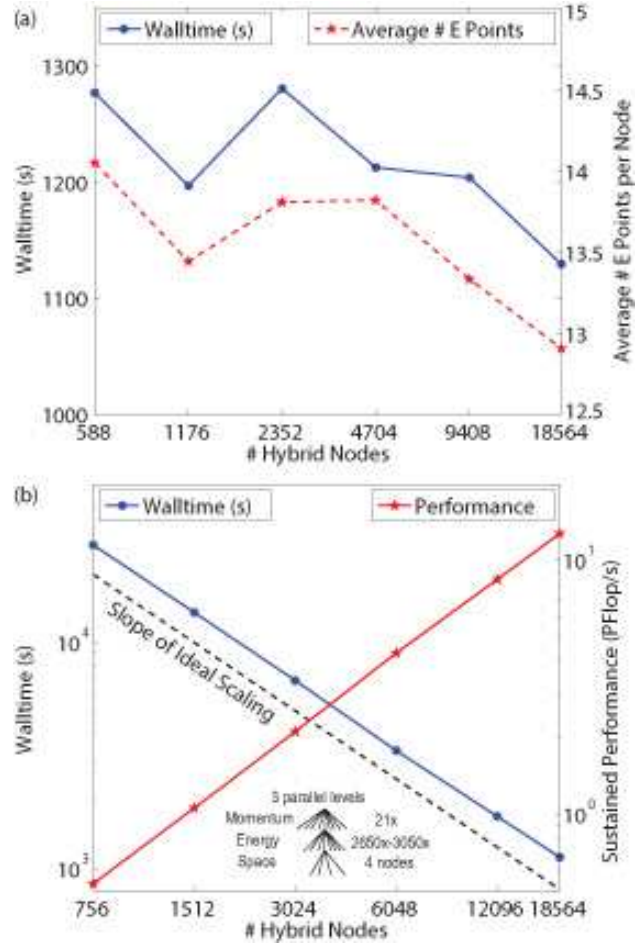}
\caption{OMEN scaling performance on Titan for a Si DG UTBFET as in
  Fig.~\ref{fig:motivation}(c) with $t_{body}$=5 nm, $L_s$=$L_c$=20
  nm, $L_g$=38.2 nm, $V_{ds}$=0.6 V, and $N_A$=23040 atoms. In all the
  simulations, one Schr\"odinger-Poisson iteration is considered, the
  total number of $k$-points is equal to 21, the 1-D spatial domain
  decomposition is performed on 4 hybrid nodes, and three levels of
  parallelization are used. (a) Weak scaling. The total number of
  energy points varies so that each node deals with 13 to 14 of
  them, regardless of the number of used nodes. Note that slight
  variations are unavoidable, as indicated by the dashed red curve,
  because the energy grid is not an input parameter, but automatically
  generated by OMEN based on the minimum and maximum allowed distance
  between two consecutive energy points. (b) Strong scaling and
  sustained performance. The number of energy points per momentum
  varies from 2650 up to 3050.}
\label{fig:scaling}
\end{figure}

\subsection*{D. Scalability}
The weak and strong scalability of OMEN on Titan, after importing the
required Hamiltonian and overlap matrices from CP2K, are presented in 
Fig.~\ref{fig:scaling} and the data summarized in Tables \ref{tab:2}
and \ref{tab:3}. A Si double-gate ultra-thin-body transistor, as in
Fig.~\ref{fig:motivation}(c), with a body thickness $t_{body}$=5 nm,
$L_s$=$L_d$=20 nm, $L_g$=38.2 nm, and a total of 23040 atoms has been
chosen as test structure of realistic dimensions \cite{doris}. In
each run, OMEN computed one Schr\"odinger-Poisson iteration for
one single bias point with 21 $k$-points and FEAST+SplitSolve on 4
hybrid nodes. Increasing the number of bias points or self-consistent
iterations does not change the scalability of the code since each
point/iteration is processed sequentially, one after the other, and
the workload is dynamically redistributed after each step \cite{parco}.

In the weak scaling experiment the average number of energy points
that each node deals with should ideally remain constant, but it
varies here between 12.9 and 14.1, as detailed in Table
\ref{tab:2}. These differences stem from the fact that the number of
calculated energy points is not a direct input parameter, but depends
on others, as explained in the caption of Fig.~\ref{fig:scaling}. If
the extracted simulation times are normalized with the average
number of energy point per node, as in the fourth column of Table
\ref{tab:2}, it can be seen that the weak scaling efficiency is good
with $\sim$5\% variation across all the nodes.

Figure \ref{fig:scaling}(b) shows the strong scaling results of
OMEN. The total number of energy points is the same in all the
runs (59908), but it varies with the momentum, i.e. $E$ depends on
$k$. The simulation time decreases almost linearly when scaling from 756 
up to 18564 nodes, reaching a parallel efficiency of 97\%. This is not
a surprise since the loop over the momentum and energy points is
embarrassingly parallel, but it demonstrates that the code
performs as expected.

\subsection*{E. Peak Performance}
OMEN running FEAST+SplitSolve for the same UTB transistor as before
initially reached a sustained performance of 12.8 PFlop/s in
double-precision, as indicated in Table \ref{tab:3}. The solution of
the OBCs and Eq.~(\ref{eq:wf}) for each of the 59908 computed energy
points consumes 241 TFLOPs, 11 for the OBCs on the CPUs (5\%) and 230
for SplitSolve on the GPUs (95\%). The bottleneck is the LU
factorization and backward substitution in Alg.~\ref{alg:1}. They are
done with the MAGMA function \verb|zgesv_nopiv_gpu| that does not
perform as well on Titan as on Piz Daint. Replacing
\verb|zgesv_nopiv_gpu| with \verb|zhesv_nopiv_gpu| and profiting from
the property that the matrix $A=(E\cdot S-H)$ is Hermitian in 2-D
structures (UTBFET) helps decrease SplitSolve's execution time per
energy point \cite{yamazaki}. By combining this trick with further
profiling and tuning of the code as well as algorithm adaptations to
Titan, the double-precision sustained performance increased to 15.01
PFlop/s in production mode. Exactly the same structure as the one
producing 12.8 PFlop/s was used for that purpose, but the simulation
time decreased from 1130 down to 912.5 sec, while the number of
operations per energy point went down from 241 to 228 TFLOPS. 

\begin{table}
\centering
\begin{tabular}{|c|c|c|c|}
\hline
Hybrid Nodes&Time (s)&Avg.~$E$/node&Avg.~Time/$E$ (s)\\
\hline
588&1277&14.1&90.8\\
\hline
1176&1197&13.4&89\\
\hline
2352&1281&13.8&92.7\\
\hline
4704&1213&13.8&87.7\\
\hline
9408&1204&13.3&90.3\\
\hline
18564&1130&12.9&87.5\\
\hline
\end{tabular}
\caption{Weak scaling data corresponding to Fig.~\ref{fig:scaling}(a).}
\label{tab:2}
\end{table}

\begin{table}
\centering
\begin{tabular}{|c|c|c|c|}
\hline
Hybrid Nodes&Time (s)&$||$ Eff. (\%)&PFlop/s\\
\hline
756&26975&100&0.54\\
\hline
1512&13593&99.2&1.06\\
\hline
3024&6806&99.1&2.12\\
\hline
6048&3415&98.7&4.23\\
\hline
12096&1711&98.5&8.45\\
\hline
18564&1130&97.3&12.8\\
\hline
\hline
18564&912.5&-&15.01\\
\hline
\end{tabular}
\caption{Strong scaling data corresponding to
  Fig.~\ref{fig:scaling}(b) and to the run that reached 15 PFlop/s
  (last line).}
\label{tab:3}
\end{table}

\begin{figure}
\centering
\includegraphics[width=0.5\linewidth]{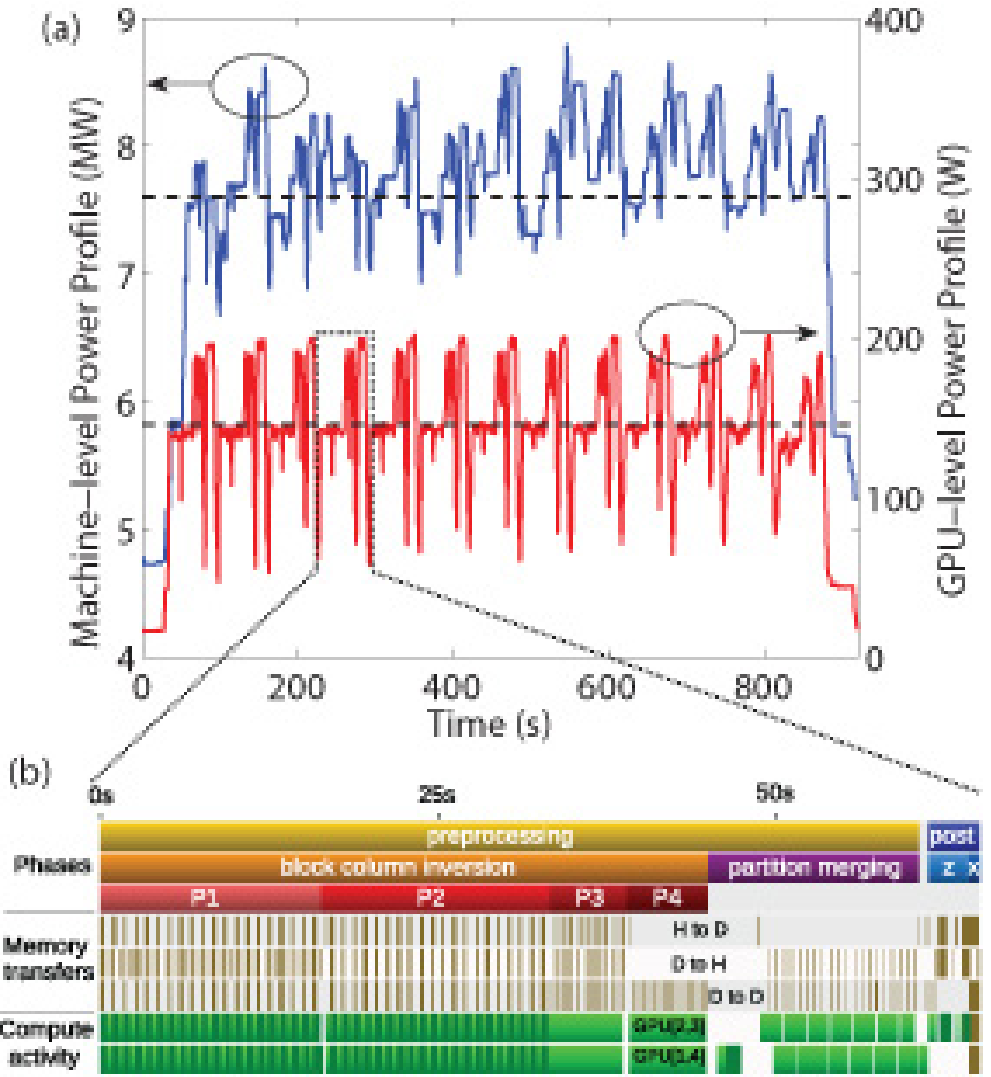}
\caption{(a) Machine-level (left $y$-axis, blue line) and GPU-level
  (right $y$-axis, red line) power profile extracted from the run that
  reached a sustained performance of 15 PFlop/s. On average, Titan
  consumed 7.6 MW during this UTB simulation (1975 MFLOPS/W), whereas
  each GPU utilized 146 W (5396 MFLOPS/W). The machine-level power
  profile includes the hardware usage (CPU+GPU), the pumping power
  used by the XDPs, the fan energy from the blower in each cabinet, as
  well as the line loss from the switchboards to the cabinets. (b)
  Typical GPU activity measured with nvprof during the
  calculation of one energy point (4 GPUs with indices 1 to 4 are 
  involved). The SplitSolve phases from Fig.~\ref{fig:pp} are
  indicated together with the host-to-device (H-to-D), device-to-host
  (D-to-H), and device-to-device (D-to-D) memory transfers as well as
  the compute activity.}
\label{fig:power}
\end{figure}

The machine- and GPU-level power profiles of the simulation that
reached 15 PFlop/s are plotted in Fig.~\ref{fig:power}(a). The 13
energy points that each group of 4 GPUs treats can be identified at
both levels, but the GPU-level matches better with the different
algorithm phases. The peak power consumption of Titan during this run is 
equal to 8.8 MW, its average to 7.6 MW, which corresponds to 1975
MFLOPS/W. At the GPU level, the computational efficiency increases to
5396 MFLOPS/W. Finally, in Fig.~\ref{fig:power}(b), the typical GPU
utilization (measured with \verb|nvprof|) is reported for the
calculation of one energy point. Note the high utilization and
well-overlapped computation-communication pattern.

\section*{\raggedright \Large \bf 6. CONCLUSION}\label{sec:conclusion}
By linking two existing applications, CP2K and OMEN, developing
SplitSolve, an innovative, GPU-based, algorithm, and combining it
with a parallel eigenvalue solver, FEAST, we have been able to
redefine the limit of \textit{ab-initio} quantum transport
simulations. Significant improvements have been achieved in terms of
structure dimensions (10$\times$ larger), sustained performance (15
PFlop/s), and time-to-solution ($>$50$\times$ shorter). While standard
approaches are typically limited to 1000 atoms, we have reached more
than 50000 for a 3-D nanowire structure and 23040 for a 2-D
ultra-thin-body.

SplitSolve heavily relies on the structure of the matrices encountered
in quantum transport calculations (block tri-diagonal + sparse
right-hand-side) to deliver its best efficiency, but these properties can
be found in other research fields such as computational fluid dynamics or
in the solution of the Poisson equation. Hence, our multi-GPU sparse
linear solver is not limited to one single problem, rather it may be
applied to others.

From 2011 to 2015 the sustained performance of OMEN increased by one
order of magnitude, going from 1.28 PFlop/s on the former Cray-XT5
Jaguar up to 15 PFlop/s on Titan, mainly through algorithmic and
hardware improvements. A roofline analysis of SplitSolve and FEAST
shows that both algorithms have high arithmetic intensity and are
clearly compute bound. It can thus be expected that OMEN will run
efficiently on future supercomputing systems offering lower relative
memory bandwidth, but higher computational power such as the
next-generation CORAL machines.

\section*{\raggedright \Large \bf ACKNOWLEDGEMENT}
This work was supported by SNF Grant No. PP00P2\_133591, by the
Hartmann M\"uller-Fonds on ETH-Research Grant ETH-34 12-1, by the
Platform for Advanced Scientific Computing in Switzerland (ANSWERS),
by the European Research Council under Grant Agreement No 
335684-E-MOBILE, by the EU FP7 DEEPEN project, and by a grant from the
Swiss National Supercomputing Centre under Project No. s579. 
This research also used resources of the Oak Ridge Leadership
Computing Facility at the Oak Ridge National Laboratory, which is
supported by the Office of Science of the U.S. Department of Energy
under Contract No. DE-AC05-00OR22725. The authors would like to thank
Jack Wells, James Rogers, Don Maxwell, Scott Atchley, Devesh Tiwari,
and colleagues at ORNL for giving them access to Titan and for greatly
supporting their work on this machine, Ichitaro Yamazaki from UT
Knoxville for his help with the MAGMA library, and Peter Messmer from
NVIDIA for his GPU expertise. 


\bibliographystyle{IEEEtran}

\end{document}